# Indium Tin Oxide film characterization using the classical Hall effect

L.H. Willems van Beveren, E. Panchenko, N. Anachi, L. Hyde, D. Smith, T.D. James, A. Roberts and J.C. McCallum

*Abstract* — we have used the classical Hall effect to electrically characterize Indium Tin Oxide (ITO) films grown by two different techniques on silica substrates. ITO films have the unique property that they can be both electrically conducting (and to be used for a gate electrode for example) as well as optically transparent (at least in the visible part of the spectrum). In the near infrared (NIR) the transmission typically reduces. However, the light absorption can in principle be compensated by growing thinner films.

*Index Terms* — Hall effect, indium tin oxide, magnetic field measurement, semiconductor devices.

## I. Introduction

ITO films are important for optoelectronic device applications and this work investigates the difference in electronic properties (mobility, carrier density and resistivity) of an ITO film deposited by physical vapor deposition (PVD) in comparison with a film grown by sputtering. The films are analyzed at both room temperature (RT) and at liquid nitrogen ($LN_2$) temperatures in a van der Pauw configuration.

## II. Experimental techniques

### A. Wafer materials ($SiO_2$ on Si)

We have grown ITO films on 2 different $Si/SiO_2$ substrates. The sputtered ITO film (125 nm thickness) was grown on a 2" $Si/SiO_2$ wafer (*p*-type resistivity of 1-20 Ohm-cm) with a $SiO_2$ (dry oxidation) thickness of 90 nm ($1100°C$ for 4 hours in 20% $O_2$ flow). These $Si/SiO_2$ thicknesses were verified by ellipsometry. This is a very sensitive and non-invasive tool, which even allows us to observe the existence of a thin (<2 nm) native oxide.

The same $SiO_2$ dielectric was recently tested for electrical properties by capacitance-voltage (C-V) and deep level transient spectroscopy (DLTS) measurements on fabricated metal-oxide-semiconductor (MOS) capacitors. The as-grown film is not leaking current, but has a number of trap states and fixed oxide charge and does not show inversion up to 10 V.

L.H. Willems van Beveren, E. Panchenko, N. Anachi, T.D. James, A. Roberts, J.C. McCallum are with the School of Physics, University of Melbourne, VIC 3010, Australia.
L. Hyde, D. Smith are with the Melbourne Centre for Nanofabrication, VIC 3168, Australia.
Corresponding author: email:laurensw@unimelb.edu.au.

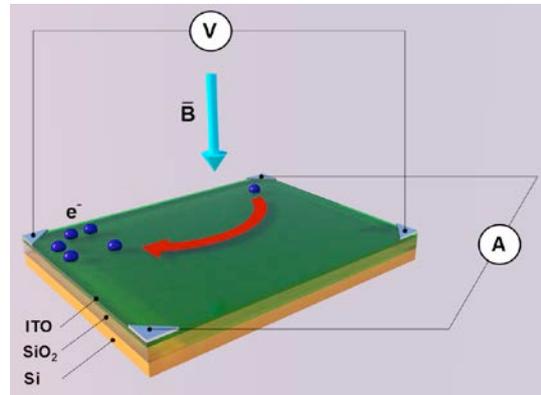

Fig. 1. Schematic representation of the Hall effect in of an ITO film (green), deposited on a silica layer, on silicon substrate (yellow). A magnetic field B deflects the electron flow (A) to one of the electrodes. Due to the excess concentration of electrons near one electrode a resulting Hall voltage (V) can be measured.

A forming gas anneal (5% $H_2$ / 95% Ar) at $450°C$ for 15 minutes significantly clears up the $Si/SiO_2$ interface and moves the flatland voltage back to approximately 0V. This demonstrates that the oxides grown at the Melbourne Centre for Nanofabrication (MCN) are adequate for our purposes, but can be optimized further by various annealing techniques.

The evaporated ITO film of 12.5 nm thickness was deposited by PVD on 4" Si wafer (*p*-type resistivity of 1-10 Ohm-cm). In this case, a 200 nm thick $SiO_2$ layer was deposited first (also using PVD) to act as an insulation layer.

### B. Sputtering

The sputtering process (in an Ar environment) for creating the 125 nm film of ITO takes place at room temperature. The composition of the ITO sputter target is $In_2O_3$: 90 wt. % $SnO_2$: 10 wt. %), and is preserved in the process. This is, in contrast to PVD deposited films, and the composition can also be experimentally obtained by Rutherford Back Scattering (RBS) spectroscopy. Optical transmission experiments (visible / NIR) will permit further quantification.

### C. Evaporation + annealing procedure

The PVD deposited film of In-Sn was grown using $InO_3$: 90 wt. % $SnO_2$: 10 wt. %) target at an evaporation rate of 0.8 A/s (the slower the evaporation rate the better the quality of the film). The resulting film was then annealed in an oxygen atmosphere to create an oxidized film of tin oxide (ITO). A noticeable optical difference can be seen by eye between as-grown and the annealed PVD film (after annealing the ITO film becomes transparent), which is attributed to a change in the electronic band gap of the material produced and confirms successful formation of ITO.

### D. Hall effect setup

We have used a Janis flow cryostat (ST-300S) with room temperature operated electro-magnets (Kepco BOP72-28) for the Hall effect (see Fig.1) analysis of the films. The DC magnetic field can be swept over a range of -0.8 to +0.8 T and the temperature can be controlled from RT down to 77 K (or even 2 K if liquid Helium is available). The resistivity and Hall voltages of the films were extracted using 4-terminal van der Pauw geometry. Here electric contacts are made to the corners of a square piece of the film to be investigated. A 6220 current source, a 2000 voltage meter and a 6487 current-meter (all Keithley) were used to measure the current that flows through the films. A 7001 matrix switch module in combination with a 7065 Hall effect card are essential to cycle through the different current-voltage configurations in order to efficiently extract the correct averaged longitudinal and Hall voltages.

## III. EXPERIMENTAL RESULTS

Using the techniques described above we have:

- Showed that the Hall resistance $R_{xy}$ vs applied *B*-field is linear for both PVD and sputtered films, which results in a Hall slope (Ohm/T) and the corresponding carrier density (see Fig. 2) once the resistivity is known.
- Found that the 125 nm thick, sputtered ITO film has a RT resistivity of 3.6 mOhm-cm. The 3D carrier density of electrons equals $n_{3D}=3.31 \times 10^{20}$/cm$^3$ and results in a mobility of $\mu_e$=5.1 cm$^2$/Vs.
- In comparison, the 12.5 nm thick annealed PVD film has a RT resistivity of 815 µOhm-cm. The 3D carrier density of electrons equals $n_{3D}=3.6 \times 10^{20}$/cm$^3$ and results in a mobility of $\mu_e$=21.3 cm$^2$/Vs.
- Both films are metallic in nature and have ~mOhm-cm RT resistivity (B=0 T), even though their thicknesses are quite different (12.5 nm vs 125 nm). Their carrier density and mobilities are slightly different (the carrier density and mobility of the PVD annealed film is greater than the sputtered film's) causing a difference in the resistivity. Note that the same difference was observed in [2].

Also, as byproducts of our experiments we noticed that:

- Wire bonding is possible on a 12.5 nm thin ITO film (PVD). This avoids the step of making Cr bonding pads in the corners of the ITO film c.f. Fig. 1.
- The resistivity of films does not change significantly at low temperatures (LN$_2$). This is similar to the behavior of a metallic system.

## IV. FUTURE DIRECTIONS

Since ITO is a conducting oxide it is possible to change its carrier concentration by gating, such as in a MOS-capacitor structure. This allows us then to characterize ITO carrier concentration on applied gate voltage. This change in density has shown to create a large change in the index of refraction (for a particular wavelength), which is extremely interesting for plasmonic device applications.

Finally, further investigations will involve a superconductivity transition in the ITO film [3] at milli-Kelvin temperature and the possibility of using these ITO films as optically transparent gate electrodes for semiconductor transistor architectures [4].

## CONCLUSION

We have fabricated ITO films by sputtering and evaporation and measured their electrical properties using the classical Hall effect. Although, the carrier densities of both films are similar, the carrier mobility of the PVD film is higher. The properties of annealed PVD films are similar to high-quality ITO films obtained by other groups [5, 6].

## ACKNOWLEDGMENT

This work was performed in part at the Melbourne Centre for Nanofabrication (MCN) in the Victorian Node of the Australian National Fabrication Facility (ANFF).

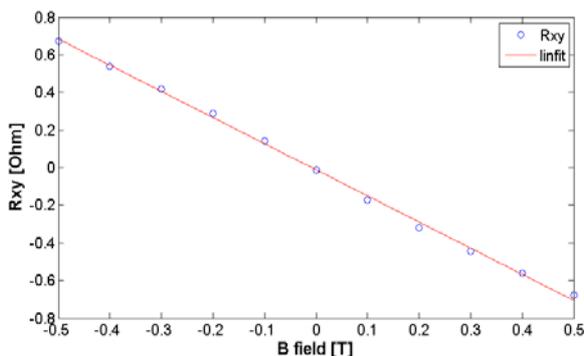

Fig. 2. Graph shows the RT magnetic field dependence of the Hall resistance for the PVD deposited ITO film. The direction of the slope is consistent with an *n*-type semiconductor (electrons are majority carriers).